%% file: main.tex
\documentclass[aps,superscriptaddress, reprint,showpacs,amsmath,amssymb,prl, groupedaddress]{revtex4-2}
\usepackage{graphicx}
\usepackage{xcolor}
\usepackage[unicode=true, breaklinks=false, backref=false, colorlinks=true, linkcolor=blue, citecolor=blue, urlcolor=blue]{hyperref}
\usepackage{physics}
\usepackage{dsfont} 
\usepackage{mathtools}

\usepackage[utf8]{inputenc}
\usepackage[normalem]{ulem}
\usepackage{amsmath}
\usepackage{amssymb}
\usepackage{xargs}
\usepackage{bigints}
\usepackage{xcolor}
\usepackage{empheq}
\usepackage{array}[=2016-10-06]
\usepackage{nicematrix}
\usepackage{enumitem}

\newcommand{\e}[1]{\mathrm{e}^{#1}}

\newcommand{\densmat}{\hat{\rho}}

\newcommand{\gL}{\gamma_L}
\newcommand{\gR}{\gamma_R}
\newcommand{\gLp}{\gamma_L^+}
\newcommand{\gLm}{\gamma_L^-}
\newcommand{\gRp}{\gamma_R^+}
\newcommand{\gRm}{\gamma_R^-}
\newcommand{\GL}{\Gamma_L}
\newcommand{\GR}{\Gamma_R}
\newcommand{\aI}{\text{Im}(\alpha)}
\newcommand{\aR}{\text{Re}(\alpha)}
\newcommand{\bR}{\text{Re}(\beta)}
\newcommand{\bI}{\text{Im}(\beta)}
\newcommand{\R}[1]{\text{Re}(#1)}
\newcommand{\I}[1]{\text{Im}(#1)}
\newcommand{\de}{\delta}
\newcommand{\sLp}{\sigma_+^{(L)}}
\newcommand{\sLm}{\sigma_-^{(L)}}
\newcommand{\sRp}{\sigma_+^{(L)}}
\newcommand{\sRm}{\sigma_-^{(L)}}
\newcommand{\gr}{g_{\text{res}}}
\newcommand{\go}{g_{\text{off}}}

\newcommand{\mygreen}[1]{\textcolor[rgb]{0, 0, 0}{#1}}
\newcommand{\myblue}[1]{\textcolor[rgb]{0, 0, 0}{#1}}

\begin{document}

\title{Transport approach to quantum state tomography}

\author{Jeanne Bourgeois$^{1,2,3}$, Gianmichele Blasi$^1$ and G\'eraldine Haack$^1$ \vspace{0.2cm}}
\affiliation{$^1$Department of Applied Physics, University of Geneva, 1211 Geneva, Switzerland \\
$^2$Institute of Physics, Ecole Polytechnique F\'ed\'erale de Lausanne (EPFL), CH-1015 Lausanne, Switzerland \\
$^3$Center for Quantum Science and Engineering, Ecole Polytechnique F\'ed\'erale de Lausanne (EPFL), CH-1015 Lausanne, Switzerland}

\date{September 15, 2025}

\begin{abstract}
Quantum state tomography (QST) is a central task for quantum information processing, enabling quantum cryptography, computation, and state certification. Traditional QST relies on projective measurements of single- and two-qubit Pauli operators, requiring the system of interest to be isolated from environmental dissipation. In this work, we demonstrate that measuring currents and associated transport quantities flowing through a quantum system in an open configuration enable the reconstruction of its quantum state. This result relies on an exact relation between transport quantities and the Krylov subspaces associated with the Lindbladian which encodes the dynamical evolution of an open quantum system. 
We illustrate this transport approach to QST with the explicit example of a two-qubit system embedded in a two-terminal setup. As a direct consequence of our framework, we are able to provide a transport-based entanglement measure to certify the presence of quantum correlations, expressing the concurrence in terms of current averages and correlations function only. Our findings are analytical, providing fundamental insights into quantum information processing in open quantum systems. They establish new connections between the fields of mesoscopic physics and quantum information theory.
\end{abstract}

\maketitle

\noindent\textit{Introduction.--} Demonstrating the presence of entanglement between multiple qubits is one of the key milestones for efficient quantum information processing, requiring the possibility to achieve quantum state tomography (QST) of multi-partite systems \cite{2004, Nielsen2012, Vogel1989}. Seminal experimental works based on different platforms have realized QST involving one to multiple qubits, with superconducting platforms \cite{Filipp2009, Bianchetti2010}, trapped ions \cite{Roos2004, Haeffner2005} and Nitrogen-vacancy centers \cite{Leskowitz2004, Wrachtrup2006}. QST protocols are typically based on linear inversion of local and global Pauli observables directly related to elements of the density operator, and on numerical optimization through various statistical estimators \cite{James2001, Bartkiewicz2016} or classical shadow algorithms \cite{Huang2020}. Remarkably, QST protocols contributed very recently to demonstrate loophole-free violation of Bell inequalities using solid-state qubits \cite{Thapliyal2016, Kurpiers2018, Storz2023}, and are starting to be explored from the point of view of novel advanced numerical techniques \cite{Torlai2018}. 

While uncontrolled dissipation leading to noise is considered as detrimental in the above experimental achievements, its role towards quantum information processing and its applications has considerably changed in the past two decades. On one hand, engineered dissipation has been shown to constitute an alternative lever for generating and manipulating quantum resources, through stabilization of entangled states in decoherence-free subspaces \cite{Lidar1998, Kwiat2000, Kielpinski2001, Kielpinski2002, Deng2007} and autonomous quantum error correction codes \cite{Reiter2017, Xu2023, Hillmann2023}. On the other hand, it remains an open question whether uncontrolled dissipation, in the form of heat from a thermodynamic point of view, can also constitute a resource for quantum information processing. As a consequence, several theoretical proposals for autonomous generation of entanglement in open quantum systems setups have followed \cite{Eisler2005, Hartmann2007, Quiroga2007, Bellomo2013, BohrBrask2015, Hewgill2018, Das2022,Khandelwal2024}. 
In particular, the successful functioning of an entanglement engine was shown to be conditioned on a minimal critical charge or heat current flowing through two interacting qubits in a two-terminal device \cite{Khandelwal2020, Farina2023, FrancescoDiotallevi2024}. This result paved the way for exploiting heat as a witness of entanglement, an approach that was generalized to quantum properties in \cite{deOliveiraJunior2025}. All these contributions suggested a connection between transport properties in an open quantum system and its quantum state.

In this work, we provide an affirmative answer to the fundamental question whether transport measurements, in the form of current averages and fluctuations, can provide a complete characterization of the state of an open quantum system. 
We derive exact relations between transport quantities - currents, time-derivatives of the currents, and  current correlation functions - and density operator's elements. We show that these relations originate from the Krylov subspaces of the populations of the system generated by the Lindbladian superoperator of the open quantum system. This provides a general theoretical framework and a fundamental understanding of the underlying principles behind a transport approach to QST. We illustrate these findings with a complete analytical description of two interacting qubits weakly coupled to two Markovian reservoirs, their dynamics being assessed within a master equation approach. We show that this transport-based tomography scheme only requires the \textit{a priori} knowledge of the system-bath coupling strengths. All other parameters, including those governing the system's unitary dynamics and local pure dephasing, can be extracted from additional transport-based quantities. Importantly, these results allow us to express the concurrence, an entanglement measure, in terms of transport quantities only. This further establishes fundamental connections between the quantum state of an open quantum system and observables measured in the environment. \\

\noindent\textit{Theoretical framework.--} We consider an open quantum system made of $N$ interacting qubits, whose internal dynamics is set by the Lindbladian $\mathcal{L}_S$. Additionally, we assume $M\leq N$ qubits to be locally coupled to Markovian thermal baths through single-particle tunneling-type interaction Hamiltonians. These reservoirs are biased in voltage and temperature such that charge and heat currents flow through the system. Assuming weak system-bath coupling compared to the typical energy scales of the multi-qubit system, the dynamics of the corresponding reduced density operator $\densmat$ is accurately captured by a local Lindblad master equation of the form \cite{Breuer2007, Schaller2014, Hofer2017, Khandelwal2020, Landi2024} ($\hbar= k_B = 1$ throughout the work):
\begin{equation}
\label{eq:Lindblad}
    \dot{\densmat} = \mathcal{L}\densmat = \mathcal{L}_S\densmat + \sum_{{j}=1}^M (\gamma_{j}^+ \mathcal{D}[\hat \sigma_+^{({j})}] +  \gamma_{j}^- \mathcal{D}[\hat \sigma_-^{({j})}] )\densmat,
\end{equation} 
where $\mathcal{D}[A] \bullet \equiv A \bullet A^\dagger - ( A^\dagger A \bullet + \bullet A^\dagger A )/2$ is the dissipator, $\hat \sigma_+^{(j)}$ ($\hat \sigma_-^{(j)}$) the raising (lowering) operators for qubit $j$ and $\gamma_{j}^{+}$ ($\gamma_{j}^-$) the rate of particles tunneling in (out) of the system from (to) reservoir $j$ (details on our analytical method are provided in the Supp. Mat. \cite{SuppMat}\nocite{globalME}).
The second term in the right-hand-side expression captures the dissipative dynamics induced by the thermal baths, including relaxation and loss of quantum coherence.

Assuming the validity of Eq.~\eqref{eq:Lindblad}, the super-operators $\mathcal{I}_{j} \bullet \equiv \gamma_{j}^+ \hat \sigma_+^{(j)} \bullet \hat \sigma_-^{(j)} - \gamma_{j}^- \hat \sigma_-^{({j})} \bullet \hat \sigma_+^{({j})}$ enable the calculation of the current operators' averages and higher-order moments, noted below with the compact notation 
$I_P(t) \equiv \Tr[\prod_{j\in P} \mathcal{I}_j \densmat(t)]$ for $P\subseteq\{1,\ldots,M\}$. 
The case $P = \{j\}$ corresponds to the averaged current from reservoir $j$ into the system, while $P= \{i, j\}$ provides the instantaneous current cross-correlations between the leads $i$ and $j$ \cite{Blasi2024}.  
The $k$-$th$ time derivative of the current moment $I_P^{(k)}(t)$ is given by 
\begin{equation}
\label{eq:current_moment}
I_P^{(k)}(t) = \Tr[\prod_{j\in P} \mathcal{I}_j \, \mathcal{L}^k \densmat(t)]\,.
\end{equation} 

Of importance for our main result formulated below, the time evolution of any initial state $\densmat_0$ is governed by the propagator $\e{\mathcal{L}t}$, such that $\densmat(t) = \e{\mathcal{L}t} \densmat_0$. This implies that the state $\densmat(t)$ remains confined to the subspace generated by successive applications of the Lindbladian to the initial state, namely the set $\{ \mathcal{L}^k \densmat_0\}_{k=0}^{\infty}$,  referred to as the \emph{Krylov space} associated with the Lindbladian $\mathcal{L}$ and the state $\densmat_0$ \cite{NANDY20251}.
Equivalently, in the Heisenberg picture the time evolution of an operator $\hat u_0$ is confined to the Krylov subspace spanned by the set of operators $\{\mathcal{L}^{\dagger k} \hat u_0\}_{k=0}^{\infty}$. 
In practice, an orthonormal basis for the Krylov space can be constructed using the Arnoldi algorithm~\cite{arnoldi,Krylov_arnoldi}: after $K$ iterations, the algorithm generates orthonormal vectors that span $\{ \mathcal{L}^k \densmat_0\}_{k=0}^{K-1}$ in the Schrödinger picture, or $\{ \mathcal{L}^{\dagger k} \hat u_0\}_{k=0}^{K-1}$ in the Heisenberg picture, providing a reduced $K$-dimensional subspace in which the time evolution can be efficiently computed.\\

\noindent\textit{Main result.--} The fundamental result behind our proposal for a transport approach to QST lies in the connection between transport quantities and the elements of the system's density operator $\densmat$ projected onto suitable Krylov subspaces. This connection arises when considering Krylov spaces generated by the operators:
\begin{equation}
\hat n_{P}\;=\; \prod_{j\in P}\hat n_{j},
\qquad P\subseteq\{1,\ldots,M\},
\end{equation}
where $\hat n_{j}$ is the occupation--number projector of qubit $j$ coupled to a thermal reservoir. The resulting sum of Krylov subspaces can be written as
\begin{equation}
\label{eq:K}
\mathcal{K}=\sum_{P\subseteq\{1,\ldots,M\}}\mathcal{K}_{P},
\quad
\mathcal{K}_{P}=\operatorname{span}\bigl\{\mathcal{L}^{\dagger k}\,\hat n_{P}\,\bigl|\,k\in\mathbb{N}\bigr\}.
\end{equation}
Here the sum runs over all subsets $P$ of qubit-reservoir indices, and each $\mathcal K_{P}$ contains the repeated actions of $\mathcal L^{\dagger}$ on the corresponding projector $\hat n_{P}$.

Elements of the density operator of the system are obtained by projecting it onto the vectors of the Krylov spaces:
\begin{equation}
\label{eq:decomp}
    p_{P, k} \equiv \langle \mathcal{L}^{k\dagger} \hat n_P , \densmat(t) \rangle = \Tr[\hat{n}_{P} \mathcal{L}^k \densmat(t)]\,,
\end{equation}
where the notation $\langle \cdot,\cdot \rangle$ corresponds to Hilbert-Schmidt inner product of two operators $\hat{A}$ and $\hat{B}$, $\langle \hat A, \hat B \rangle = \Tr[\hat A ^\dagger \hat B]$. Physically, these elements correspond to the occupation numbers $p_{P,0}(t)=\langle \hat n_P \rangle (t) \equiv n_P(t)$ ($k=0$), and their high-order time-derivatives $p_{P,k}(t) = n_P^{(k)}(t)$ ($k\geq1$).

\begin{figure*}[t]
\begin{center}
\includegraphics[width=0.8\textwidth]{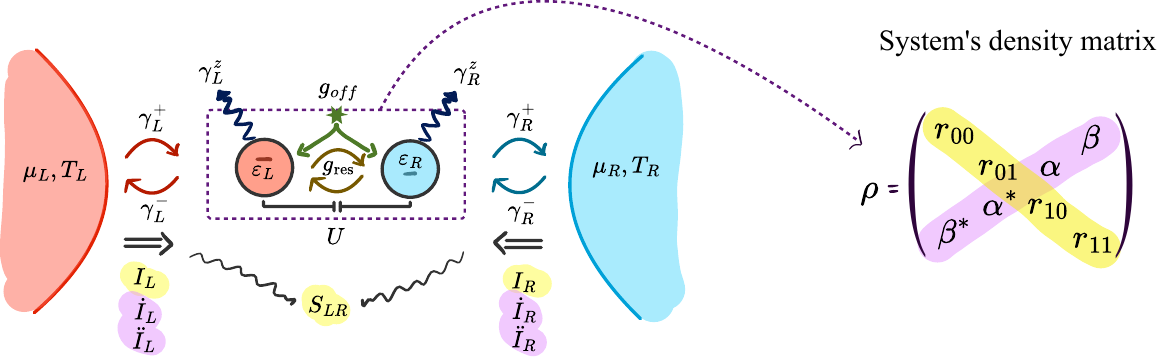}
\caption{
Illustration of the tomography of a two-qubit open quantum system through the measurement of transport observables. The represented evolution corresponds to Eq.~\eqref{eq:Lindblad_TQS}.}
\label{fig:proposal}
\end{center} 
\end{figure*}

The elements $p_{P, k}$ of Eq.~\eqref{eq:decomp} are related to the $k-th$ derivative of the current moments $I_{P'(t)}$ defined in Eq.~\eqref{eq:current_moment} through the identity:
\begin{eqnarray} 
    p_{P, k} &=& \Tr[ \Big (\prod_{j \in P} \frac{\gamma_j^+ - \mathcal{I}_j}{\Gamma_j} \Big ) \mathcal{L}^k \densmat(t)] \label{eq:QST_factor}\\
    &=& \sum_{P'\subseteq P} \Big (\prod_{i \notin P', \, j \in P} \frac{\gamma_{i}^+}{\Gamma_j} \Big ) \; (-1)^{\mid P' \mid}  \; I_{P'}^{(k)}(t)\,,  \label{eq:QST_develop}  
\end{eqnarray}
with $\Gamma_{j} \equiv \gamma_{j}^+ +\gamma_{j}^-$, the index $P'$ running through the list of all subsets of $P$, $i$ through the list of indices in $P$ not in $P'$ and $j$ through all $P$. The proof of this identity, a fundamental and technical result of this work, is provided in the End Matter.

Whether this quantum tomography protocol is complete or not depends on the size of the Krylov spaces with respect to the one of the Hilbert space $\mathcal{H}$ of the system. In general, $\text{dim}(\mathcal{K}_{P}) \leq \text{dim}(\mathcal{H})$ \cite{NANDY20251}, requiring the combination of several Krylov spaces, corresponding to different occupation-number operators, to potentially achieve complete QST. We emphasize that the size of the Krylov subspaces depends on the spectral properties of the Lindbladian, as analyzed in the Supp. Mat. \cite{SuppMat}.  For the explicit example of two-qubit state tomography discussed below, we show that complete QST can be achieved by combining Krylov spaces associated to different occupation-number operators.\\

\noindent\textit{Two-qubit quantum state tomography.--} We illustrate the transport approach to QST with the paradigmatic example of a two-qubit system coupled to two reservoirs, see Figure~\ref{fig:proposal}. We write the density matrix in the Fock basis $\{ \ket{00}, \ket{01}, \ket{10}, \ket{11}\}$ as
\begin{equation}
    \densmat = \begin{pmatrix}
r_{00} & v & x & \beta \\
v^{\ast} & r_{01} & \alpha & y \\
x^{\ast} & \alpha^{\ast} & r_{10} & z \\
\beta^{\ast} & y^{\ast} & z^{\ast} & r_{11} \\
\end{pmatrix}
\end{equation}
with the diagonal elements corresponding to populations (real numbers) and off-diagonal elements to coherences (complex numbers).
The two qubit-reservoir pairs are referred to as “left” (L) and “right” (R).  

Without any prior knowledge of the internal evolution of the two-qubit system, the application of Eq.~\eqref{eq:QST_develop} for $k=0$ to the population projectors $\hat n_L$, $\hat n_R$ and $\hat n_L \hat n_R$ allows us to determine the populations of the system via the left and right currents $I_L(t)$ and $I_R(t)$ and their instantaneous cross-correlation function $S_{LR}(t) = I_{LR}(t) - I_L(t) I_R(t)$: 
\begin{subequations}
\label{eq:populations}
\begin{align}
    r_{01}(t) &= -\frac{S_{LR}(t)}{\GL \GR} - \frac{I_R(t)-\gR^+}{\GR} \frac{I_L(t) + \gamma_L^-}{\Gamma}, \\
    r_{10}(t) &= - \frac{S_{LR}(t)}{\GL \GR} - \frac{I_L(t)-\gL^+}{\GL} \frac{I_R(t) + \gamma_R^-}{\Gamma}, \\
    r_{11}(t) &= \frac{S_{LR}(t)}{\GL \GR} + \frac{I_L(t)-\gL^+}{\GL} \frac{I_R(t)-\gR^+}{\GR},
\end{align}
\end{subequations}
with $r_{00}(t)$ being reconstructed from the trace property $\Tr{\densmat(t)}=1$. Details of this derivation are provided in the End Matter. The set of equations \eqref{eq:populations} attests that average currents and cross-correlated noise are sufficient to determine the populations of a two-qubit system at any given time, given that the couplings  $\gamma_{L/R}^{\pm}$ are known \emph{a priori}. This assumption is consistent with state-of-the-art experimental setups, as discussed in the conclusion of this work. 
Interestingly, although the presence of baths influences the evolution of all elements of the density matrix, the current quantities $I_L(t)$, $I_R(t)$ and $S_{LR}(t)$ at a fixed time $t$ depend solely on the populations $r_{ij}(t)$ and not on the coherences.

To access the coherences, we make use of Eqs.~\eqref{eq:QST_factor} and ~\eqref{eq:QST_develop} with $k=1$ and $k=2$. The corresponding equations then depend on the system's internal evolution. In the following, we consider it to be given by the Lindbladian
\begin{align} 
\label{eq:Lindblad_TQS}
    \mathcal{L}_S \bullet &= - i [\hat H_S, \bullet] 
     + \sum_{{j}=L,R} \frac{\gamma_j^z}{2} \mathcal{D}[\hat \sigma_z^{({j})}] \bullet
\end{align}
with $H_S$ a generic two-qubit Hamiltonian that induces a coupled dynamics for the diagonal and off-diagonal elements of $\densmat$,
\begin{align}
    \label{eq:Hs}
    \hat{H}_S &= \varepsilon_L \hat{n}_L + \varepsilon_R \hat{n}_R + U \hat{n}_L \hat{n}_R  \nonumber \\
    &\qquad + (g_\text{res} \, \hat\sigma_+^{(L)} \hat\sigma_-^{(R)} + g_\text{off} \, \hat\sigma_+^{(L)} \hat\sigma_+^{(R)}) + h.c.\,.
\end{align}
The energies $\varepsilon_{L/R}$ set the bare energies of the qubits $L/R$, the energy $U$ captures on-site interactions (like Coulomb interactions for charge qubits), while the energies $g_\text{res}$ and $g_\text{off}$ set resonant (particle-conserving)
and off-resonant (particle-non-conserving) interactions respectively and are supposed real without loss of generality. The dissipative terms of the Lindbladian $\mathcal{L}_S$ \eqref{eq:Lindblad_TQS}, involving the z-Pauli jump operators $\hat \sigma_z^{({\alpha})}$ and proportional to rates $\gamma_z^{(\alpha)}$, capture pure dephasing processes contributing to the finite coherence time of the qubits.

\begin{table}[t]
\label{table}
\begin{NiceTabular}{c||c|c|c:c}[corners, hlines]
Case
&\Block[c]{}{Populations} 
& \Block[c]{}{Transport\\ quantities} 
& \Block[c]{1-2}{Transport\\ parameters} \\
\Hline \Hline

& $r_{00}, r_{01}, r_{10}$                                         
& \Block[c]{}{\myblue{$I_L$}, \myblue{$I_R$}, \\\myblue{$S_{LR}$}}
& \Block[c]{1-2}{\mygreen{$\gamma_L^{\pm}$}, \mygreen{$\gamma_R^{\pm}$}} \
\\ 

\Block[]{1-5}{} \\

Case                                                       
& \Block[c]{}{Generated\\ coherences} 
& \Block[c]{}{Additional \\ transport\\ quantities} 
& \Block[c]{1-2}{Hamiltonian param. \\ \& add. measurement \\ when $\Gamma_z$ unkown}
\\ \Hline \Hline

& $\alpha$, $\beta$                                              
& \Block[c]{}{\myblue{$\dot I_L, \ddot I_L$}\\ and\\ \myblue{$\dot I_R, \ddot I_R$}} 
& \Block[c]{}{\mygreen{$g_\text{res}$}, \mygreen{$g_\text{off}$},\\ \mygreen{$\delta$}, \mygreen{$E$}}
& \Block[c]{}{\mygreen{$I_L^{(3)}$} or \mygreen{$I_R^{(3)}$}}
\\ 

\Block[c]{}{$\delta=0$\\ $E=0$} 
& $\aI, \bI$
& \Block[c]{}{\myblue{$\dot I_L$}\\ and\\ \myblue{$\dot I_R$}}
& \Block[c]{}{\mygreen{$g_\text{res}$}, \mygreen{$g_\text{off}$}}
& \Block[c]{}{$\varnothing$}
\\ 

$g_\text{off}=0$
& $\aI, \aR$
& \Block[c]{}{\myblue{$\dot I_L, \ddot I_L$}\\ or\\ \myblue{$\dot I_R, \ddot I_R$}}
& \Block[c]{}{\mygreen{$g_\text{res}$},\\ \mygreen{$\delta$}}
& \Block[c]{}{\mygreen{$I_L^{(3)}$} or \mygreen{$I_R^{(3)}$}}
\\ 
\end{NiceTabular}
\caption{Table summarizing which transport quantities allow for the reconstruction of the populations and the coherences of the two-qubit system. The rightmost column lists the transport and Hamiltonian parameters which \textit{a priori} knowledge is required to achieve QST. If certain parameters are not known (for example here, dephasing rate $\Gamma_z=\gamma_L^z + \gamma_R^z$), we list additional transport quantities that can be exploited to assess them (see Supp. Mat. for details). 
The general case (no specific assumption) corresponds to Eq.~\eqref{eq:Hs}.}
\label{tab:coh}
\end{table}

Given this internal evolution and applying Eq.~\eqref{eq:QST_develop} for $P=\{L\},\{R\}$ and $k=1,2$, the first time-derivatives of the left and right currents, $\dot I_L(t)$ and $\dot I_R(t)$, determine the imaginary part of the coherences $\alpha$ and $\beta$, 
\begin{subequations}
\label{eq:cohI}
\begin{align}
    -4 \, g_\text{res} \, \textrm{Im} \,\alpha (t) &= \frac{\Dot{I}_L(t)}{\GL} + I_L(t) -\frac{\Dot{I}_R(t)}{\GR} -I_R(t), \label{eq:alphaI} \\
    -4 \, g_\text{off} \, \textrm{Im} \,\beta (t) &= \frac{\Dot{I}_L(t)}{\GL} + I_L(t) +\frac{\Dot{I}_R(t)}{\GR} +I_R(t) \label{eq:betaI}
\end{align}
\end{subequations}
while further measuring their second time-derivatives, $\ddot I_L(t)$ and $\ddot I_R(t)$, yields the real part of $\alpha$ and $\beta$ (see End Matter for details). 

Higher time-derivatives of the currents, as well as time-derivatives of the cross-correlation function, only add redundant information on the density matrix. In other words, $\mathcal L^{\dagger k}\,\hat n_j$ for $j=L,R$ and $k=0,1,3$, in addition to $\hat n_L \hat n_R$, generate the full subspace $\mathcal K$ defined in Eq.~\eqref{eq:K}. 
Interestingly, the set of accessible coherences depends on the evolution properties of the system, and exactly corresponds to the coherences being non-zero in the steady-state regime. In Table \ref{tab:coh}, we summarize the features of the transport approach to two-qubit QST for specific cases of the Hamiltonian $H_S$ in Eq.~\eqref{eq:Hs}.
Additional transport measurements may nonetheless be exploited to prone the system's evolution (see Supp. Mat. \cite{SuppMat}). 

More generally, accessing all density matrix's elements is possible, but it requires a Hamiltonian involving additional operators. In the Supp. Mat., we provide a demonstration of this generalization towards complete QST \cite{SuppMat}. \\

\textit{Steady-state regime.--} This transport approach to QST is particularly advantageous in the steady state ($ss$), where all time derivatives vanish. For all $P\subseteq \{1,...M\}$, the identity given by Eq.~\eqref{eq:QST_develop} then reduces to:
\begin{subequations}
\begin{eqnarray}
    p_{P, 0}^{(ss)} &=& \sum_{P'\subseteq P} \Big (\prod_{i \notin P', \, j \in P} \frac{\gamma_{i}^+}{\Gamma_j} \Big ) \; (-1)^{\mid P' \mid}  \; I_{P'}^{(ss)}\,,  \\
    p_{P, k}^{(ss)} &=& 0 \quad \text{ for } k\geq1.
\end{eqnarray} 
\end{subequations}
In the illustrative case of the two-qubit system, the populations and coherences are simply accessed by the measurements of steady-state currents and noise, as shown in the End Matter. \\

\textit{Transport-based entanglement measures.--}Remarkably, our demonstration of complete transport-based QST allows us to express entanglement measures in terms of currents and associated quantities only. While the concurrence is defined from the Schmidt coefficients \cite{Wootters1998}, it takes a particular simple form for X-shaped density matrices of two-qubit systems \cite{Yu2007}: $\mathcal{C} = 2 \max\{0, \vert \alpha \vert -
\sqrt{r_{00} r_{11}}, \vert \beta \vert -
\sqrt{r_{01} r_{10}}\}$. 
In the illustrative case of two energy-degenerate qubits ($\delta = 0$) with resonant interaction only ($g_\text{off} = 0$), initialized in their ground state, the density matrix $\rho(t)$ remains X-shaped at all time $t$ and its concurrence can then be expressed as:
\begin{widetext}
\begin{align}
\label{eq:ent}
    \mathcal{C} =  \max \Big \{0, & \abs{\frac{1}{g_\text{res}} \Big (\frac{\dot{I}_L}{\Gamma_L} + I_L \Big)}
    - 2
    \sqrt{
    \Big (\frac{S_{LR}}{\GL\GR} + \frac{I_L - \gLp}{\GL} \frac{I_R - \gRp}{\GR} \Big ) 
    \Big (\frac{S_{LR}}{\GL\GR} + \frac{I_L + \gLm}{\GL} \frac{I_R + \gRm}{\GR} \Big ) } \; \Big \}.
\end{align}
\end{widetext}
Equation \eqref{eq:ent} shows that certifying the presence of entanglement in an open quantum system does not require QST through decoupling the system form its environments and performing projective measurements. In the Supp. Mat. \cite{SuppMat}, we provide the expression of the concurrence $\mathcal{C}$ in the general case. These results formally connect entanglement measures of a quantum system and transport quantities accessible in an open device configuration. \\

\noindent \textit{Experimental relevance and conclusion.--} 
Transport-based QST relies on the measurement of time-resolved current traces, particle or charge currents. For emblematic quantum transport setups with solid-state qubits like quantum dots, current traces correspond to projective measurements of incoming charges into metallic contacts. Repetition of these measurements enable access to averaged time-resolved current traces of interest to our approach, see the reviews \cite{Kouwenhoven1997, vanderWiel2002}.

Another experimental aspect that is central to our approach concerns the knowledge of the system dynamics and the system–bath coupling strengths. Although our method relies on this prior information, this does not represent a limitation: the coupling strengths can be accurately determined with state-of-the-art spectroscopy measurements in solid-state platforms \cite{Liu2014, Stockklauser2017, Wang2023, Duprez2024}, while recent experiments on hybrid setups have demonstrated that the system dynamics and possible dephasing can also be estimated. Crucially, we show in the Supp. Mat.~\cite{SuppMat} that our transport-based approach itself provides access to these parameters: by measuring current derivatives at different times, one can directly infer the system’s internal dynamics and dissipation rates. Therefore, such a necessary knowledge does not constitute a limitation to the applicability of our proposal. 

This work is a significant step towards demonstrating complete QST of a quantum system from measuring out-of-equilibrium environments and associated transport properties. We provide a general theoretical framework based on the concept of Krylov subspaces generated by the Lindbladian to prove the validity of this approach. We illustrate it with the specific case of two qubits embedded into a two-terminal device. This work paves the way to quantum information processing in out-of-equilibrium quantum devices. This is of particular interest towards error mitigation \cite{Cramer2010} and neuromorphic computing \cite{Markovi2020, Mehonic2024, Kudithipudi2025}.\\

\noindent \emph{Acknowledgements.--} We are thankful for fruitful discussions with Landry Bretheau, Gwendal F\`eve and Matteo Seclì, and useful feedbacks from Vincenzo Savona. All authors acknowledge the support of NCCR SwissMAP.

\bibliography{references.bib}

\newpage
\appendix
\section{End matter}
\label{sct:end_matter}

\subsection{Proof of Eqs.~\eqref{eq:QST_factor} and \eqref{eq:QST_develop}}

We proceed by inductance to demonstrate the main identity connecting the current superoperator and
the occupation-number operators, Eq.~\eqref{eq:QST_factor} in the main text. 

Let us first consider a single qubit connected to a single reservoir, assuming the validity of the definition of the current superoperator defined below Eq.~\eqref{eq:Lindblad}. We can therefore write for any operator $\hat{u}$ :
\begin{align}
    \Tr[\mathcal{I}_{j} \hat u] 
    &= \Tr[\gamma_{j}^+ \hat \sigma_{j}^+ \hat u \hat \sigma_{j}^- - \gamma_{j}^- \hat \sigma_{j}^- \hat u \hat \sigma_{j}^+ ] \nonumber \\
    &= \gamma_{j}^+ \Tr[ \hat \sigma_{j}^- \hat \sigma_{j}^+ \hat u]    - \gamma_{j}^- \Tr[\hat \sigma_{j}^+ \hat \sigma_{j}^- \hat u ] \nonumber \\
    &= \Tr[(\gamma_{j}^+ - \Gamma_{j} \hat n_{j}) \hat u]\,,
\end{align}
where we have used the properties of linearity and permutation under the trace, as well as the commutation relations $\{\hat \sigma_{j}^+, \hat \sigma_{j}^- \}=\mathbb{I}$. Inverting this relation, we obtain:
\begin{equation} \label{eq:single_qubit}
    \Tr[\hat n_{j} \hat u ] = \Tr[(\frac{\gamma_{j}^+-\mathcal{I}_{j}}{\Gamma_{j}}) \; \hat u ].
\end{equation}
For $m$ distinct qubits coupled to $m$ distinct reservoirs (defining the set $P$), we assume the validity of the above relation:
\begin{equation} \label{eq:single_qubit}
    \Tr[\hat n_{P} \hat u ] = \Tr[\prod_{j\in P}(\frac{\gamma_{j}^+-\mathcal{I}_{j}}{\Gamma_{j}}) \; \hat u ].
\end{equation}
For $m+1$ qubits, we add an additional qubit labeled $j_0$, that does not belong to the set $P$. Therefore, $[\hat{n}_{j_0}, \hat{n}_P]=0$ and $[\sigma_\pm^{j_0}, \hat{n}_P]=0$. This allows to write:
\begin{align}
    \Tr[\hat n_{j_0} \hat n_P \hat u] = \Tr[(\frac{\gamma_{j_0}^+-\mathcal{I}_{j_0}}{\Gamma_{j_0}}) \; \hat n_P \hat u] = \Tr[\hat n_P (\frac{\gamma_{j_0}^+-\mathcal{I}_{j_0}}{\Gamma_{j_0}}) \hat u],
\end{align}
Applying the iteration hypothesis to the operator $(\frac{\gamma_{j_0}^+-\mathcal{I}_{j_0}}{\Gamma_{j_0}}) \hat u$, we obtain for the set $P'= P\cup \{j_0\}$:
\begin{align}
    \Tr[\hat n_{P'} \hat u] &= \Tr[\prod_{j\in P'} (\frac{\gamma_j^+-\mathcal{I}_j}{\Gamma_j})  \; \hat u]\,.
\end{align}
Finally, taking the $k-th$ derivative and considering $\hat{u} = \densmat$ lead to Eq.~\eqref{eq:QST_factor}. Equation \eqref{eq:QST_develop} is obtained after a simple algebraic manipulation by inserting the definition of the current moment $I_P^{(k)}$ given by Eq.~\eqref{eq:current_moment} into Eq.~\eqref{eq:QST_factor}. This ends the proof of these two main equations establishing the connection between Krylov subspaces generated by the Lindbladian and transport quantities.

\subsection{Detailed equations for the two-qubit paradigmatic example}

In the case of a two-qubit system, Eq.~\eqref{eq:QST_develop} applies for $P=\{L\}, \, \{R\},\, \{L,R\}$, yielding
\begin{subequations}
\begin{align}
    p_{L, k}(t) =& \frac{1}{\GL}(\delta_{k,0} \gLp - I_L^{(k)}(t)) \\
    p_{R, k}(t) =& \frac{1}{\GR} (\delta_{k,0} \gRp - I_R^{(k)}(t)) \\
    p_{LR, k}(t) =& \frac{1}{\GL\GR} (\delta_{k,0} \gLp \gRp - \gRp I_L^{(k)}(t) \notag \\
    & \qquad - \gLp I_R^{(k)}(t) + I_{LR}^{(k)}(t)).
\end{align}
\end{subequations}

\noindent For $k=0$, these elements correspond to occupation numbers of the qubit states. Recalling that the number operators $\hat n_j$ act as the projector $\ketbra{1}{1}$ on qubit $j$ and as the identity elsewhere, we have 
\begin{subequations}
\begin{align}
    p_{L,0}&=\langle \hat{n}_L \rangle \equiv n_L = r_{10} + r_{11} \\
    p_{R,0}&=\langle \hat{n}_R \rangle \equiv n_R = r_{01} + r_{11}
\end{align}
\end{subequations}
the occupation numbers of the left and right qubits respectively, and 
\begin{equation}
    p_{LR,0}= \langle \hat n_L \hat n_R \rangle = r_{11}
\end{equation}
the occupation number of the doubly-occupied state. 
By introducing the current cross-correlation function $S_{LR}(t)=I_{LR}(t) - I_L(t)I_R(t)$, the above equations can be rewritten as:
\begin{subequations}
\begin{align}
    &n_L(t) = r_{10}(t) + r_{11}(t) = \frac{\gLp - I_L(t)}{\GL}, \\
    &n_R(t) = r_{01}(t) + r_{11}(t) = \frac{\gRp - I_R(t)}{\GR}, \\
    &r_{11}(t) = \frac{S_{LR}(t)}{\GL\GR} + \frac{\gLp - I_L(t)}{\GL} \frac{\gRp - I_R(t)}{\GR}.
\end{align}
\end{subequations}
A straightforward calculation leads to the set of  Eq.~\eqref{eq:populations} for the populations of $\densmat$, .

\noindent For $k\geq1$, the $p_{P,k}$ correspond to the $k$-$th$ time derivatives of the $p_{P,k}$, such that $p_{L,k}(t)=n_L^{(k)}(t)$, $p_{R,k}(t)=n_R^{(k)}(t)$ and $p_{LR,k}(t) = r_{11}^{(k)}(t)$. The evolution of the two-qubit system being given by Eqs.~\eqref{eq:Lindblad_TQS} and \eqref{eq:Hs}, we have
\begin{subequations}
\begin{align}
    \dot n_L(t) &= \gLp - \GL n_L + 2 g_{\text{res}} \aI + 2 g_{\text{off}} \bI, \\
    \dot n_R(t) &= \gRp - \GR n_R - 2 g_{\text{res}} \aI + 2 g_{\text{off}} \bI.
\end{align}
\end{subequations}
Simple algebra then yields the expressions of the imaginary parts of the off-diagonal coherences, $\text{Im} (\alpha)(t)$ and $\text{Im} (\beta)(t)$, Eq.~\eqref{eq:cohI} in the main text. 

Similarly, we have
\begin{subequations}
\begin{align}
    \ddot n_L(t) &= -\GL(\gLp - \GL n_L) \notag \\
    & \qquad+ 2 \gr^2 (n_R - n_L) + 2 \go^2 (r_{00}-r_{11}) \notag \\
    & \qquad - (\frac{\tilde \Gamma}{2} + \GL) (2 g_{\text{res}} \aI + 2 g_{\text{off}} \bI) \notag \\
    & \qquad + 2 \gr \delta \aR + 2 \go E \bR, \\
    \ddot n_R(t) &= -\GR(\gRp - \GR n_R) \notag \\
    & \qquad - 2 \gr^2 (n_R - n_L) + 2 \go^2 (r_{00}-r_{11}) \notag \\
    & \qquad - (\frac{\tilde \Gamma}{2} + \GL) (-2 g_{\text{res}} \aI + 2 g_{\text{off}} \bI) \notag \\
    & \qquad - 2 \gr \delta \aR + 2 \go E \bR,
\end{align}
\end{subequations}
with $\delta = \varepsilon_L - \varepsilon_R$ the energy detuning between the qubits, $E=\varepsilon_L + \varepsilon_R + U$ the energy of the doubly-occupied state $\lvert 11\rangle$, $\tilde{\Gamma} = \Gamma_L + \Gamma_R + 2 \Gamma_z$ the total dissipation strength and $\Gamma_z = \gamma_L^z+\gamma_R^z$ the total pure dephasing strength. After some simplifications, Eq.~\eqref{eq:QST_factor} for $j=L,R$ and $k=2$ provides the transport-based QST expressions for the real parts of the off-diagonal coherences:
\begin{widetext}
\begin{subequations} 
\label{eq:cohR}
\begin{align}
    -4 \, g_\text{res} \, \de \, \aR &= 
    \Big(\frac{\Ddot{I}_L}{\GL} + \dot{I}_L - 
    \frac{\Ddot{I}_R}{\GR} - \dot{I}_R \Big) + 
    \frac{\Tilde{\Gamma}}{2}\Big (\frac{\dot{I}_L}{\GL} + I_L - \frac{\dot{I}_R}{\GR} - I_R \Big) + 
    4g_\text{res}^2 \Big (\frac{I_L - \gLp}{\GL} - \frac{I_R - \gRp}{\GR} \Big ), \label{eq:ddI1} \\
    -4 \, g_\text{off}\, E \,\bR &= \Big (
    \frac{\Ddot{I}_L}{\GL} + \dot{I}_L + 
    \frac{\Ddot{I}_R}{\GR} + \dot{I}_R \Big) + 
    \frac{\Tilde{\Gamma}}{2}\Big (\frac{\dot{I}_L}{\GL} + I_L + \frac{\dot{I}_R}{\GR} + I_R \Big) + 
    4g_\text{off}^2 \Big (\frac{I_L - \gLp}{\GL} + \frac{I_R - \gRp}{\GR} +1 \Big)\,. \label{eq:ddI2}
\end{align}
\end{subequations}
\end{widetext}

\subsection{Steady-state transport-based QST for two qubits}

\noindent In the case of the two-qubit system, considering a zero off-resonant inter-qubit interaction ($g_\text{off}=0$), of specific interest in the context of entanglement generation from out-of-equilibrium currents \cite{BohrBrask2015, Khandelwal2020}, current conservation implies $I_L^{ss} = - I_R^{ss}\equiv I^{ss}$ and only the $\alpha$-coherence remains finite (see also \cite{Bourgeois2024}). Its real and imaginary parts can be reconstructed through the equations:
\begin{subequations}
\begin{align}
    2g_\text{res}\aI^{ss} &= - I^{ss}, \label{eq:ss_current_coh}\\
    2g_\text{res}\de \aR^{ss} &= 2g_\text{res}^2 \Big (\frac{\gLp}{\GL} - \frac{\gRp}{\GR} \Big ) - \Big (\frac{\tilde \Gamma}{2} + \frac{4 g_\text{res}^2}{\GL \GR} \frac{\Gamma}{2} \Big ) I^{ss}. \label{eq:det_gamma_z}
\end{align}
\end{subequations}
Interestingly, the first equation corresponds to previously-established connection between steady-state currents and non-zero coherences in this setup. While it was derived analytically in the steady-state regime in previous works \cite{Khandelwal2020, FrancescoDiotallevi2024, Bourgeois2024}, this work provides a transport interpretation of this connection, bringing a complete and fundamental reason to its existence (see Supp. Mat. for further insights on intra-system transport \cite{SuppMat}). Additionally, the measurement of the steady-state current averages and correlations allows us to determine the pure dephasing strength, $\Gamma_z = \frac{1}{2}(\tilde{\Gamma}- \Gamma)$ with $\Gamma = \GL + \GR$, if the other dynamics parameters are known. Indeed, taking the time derivative of Eq.~\eqref{eq:ddI2} and its steady-state limit yields
\begin{align}
    \Big(\Big(\frac{\tilde \Gamma}{2}\Big)^2 + \frac{g_\text{res}^2 \Gamma \, \tilde \Gamma}{\GL \GR} + \delta^2\Big) I^{ss} = 2g_\text{res}^2 \tilde \Gamma \Big(\frac{\gLp}{\GL} - \frac{\gRp}{\GR}\Big)\,,
\end{align}
from which $\tilde{\Gamma}$ can be extracted. \\

%%%%%%%%%%%%%%%%%%%%%%%%%%%%%%%%%%%%%
\clearpage
\widetext
\appendix

\include{Supp_Mat_main}

\end{document}

%% file: Supp_Mat_main.tex
\section{Appendix A: Methodology}
\label{sct:methodology}

\subsection{Local master equation}

Our model of the system's evolution is based on the local Lindblad master equation \cite{Breuer2007, Schaller2014, Hofer2017, Khandelwal2020, Landi2024}  assessing the quantum dynamics of $N$ qubits, among which $M$ are individually tunnel-coupled to $M$ distinct thermal reservoirs at equilibrium (with $\hbar = k_B = 1$ throughout the text):
\begin{equation} \label{eq:Lindblad}
    \dot{\densmat} = \mathcal{L}\densmat = \mathcal{L}_S \densmat+ \sum_{j=1}^M (\gamma_j^+ \mathcal{D}[\hat \sigma_+^{(j)}] +  \gamma_j^- \mathcal{D}[\hat \sigma_-^{(j)}] )\densmat ,
\end{equation}
with $\mathcal{L}_S$ the Lindblad superoperator setting the evolution of the system in the absence of the $M$ reservoirs and $\mathcal{D}[\hat A] \bullet \equiv \hat A \bullet \hat A^\dagger - ( \hat A^\dagger \hat A \bullet + \bullet \hat A^\dagger \hat A \}/2$ the dissipator associated to the operator $\hat A$. 
Denoting $\gamma_j$ the bare reservoir-qubit coupling rate at side $j$ and $n_{F/B}^j(\epsilon) = (1\pm\text{e}^{(\epsilon-\mu_j)/T_j})^{-1}$ the probability distribution of bath $j$ (respectively for a fermionic or bosonic reservoir), the  in- and out-going coupling rates $\gamma_j^+$, $\gamma_j^-$ are given by 
\begin{equation}
\gamma_j^+=\gamma_jn_{F/B}^j(\epsilon_j) \quad \text{and} \quad \gamma_j^-=\gamma_j(1\mp n_{F/B}^j(\epsilon_j)),
\end{equation} 
with the upper sign for fermions and the lower one for bosons. Here, the probability distribution of the reservoir $j$ is evaluated at the bare energy $\epsilon_j$ of its associated qubit, while $T_j$ and $\mu_j$ denote the reservoir's temperature and chemical potential.

The local Lindblad master equation correctly captures the dynamics of the reduced density operator for the $N$-qubit system when the reservoir-qubit and inter-qubit couplings are small with respect to the other energy scales, and when the reservoirs can be considered as Markovian. For a two-qubit system with Hamiltonian $\hat{H}_S = \varepsilon_L \hat{n}_L + \varepsilon_R \hat{n}_R + U \hat{n}_L \hat{n}_R  + (g_\text{res} \, \hat\sigma_+^{(L)} \hat\sigma_-^{(R)} + g_\text{off} \, \hat\sigma_+^{(L)} \hat\sigma_+^{(R)} + h.c.)$, where $U$, $g_\text{res}$ and $g_\text{off}$ respectively set on-site, resonant (particle-conserving) and off-resonant (particle-non-conserving) interactions and are supposed real and positive without loss of generality, such conditions consist in \cite{Khandelwal2020, Landi2024, Blasi2024}:
\begin{equation}
    \gamma_L, \gamma_R \ll T_L, T_R, \abs{\epsilon_L -\mu_L}, \abs{\epsilon_R -\mu_R} \qquad \text{and} \qquad  g_\text{res}, g_\text{off} \lesssim \gamma_j.
\end{equation}
If the assumption of weak inter-qubit coupling is not satisfied, the global Master equation provides a more accurate description of the system's evolution, see for instance \cite{globalME}. While the relation between the measurements of current superoperators and occupation-number operators will be accordingly modified, there is no conceptual difference with the case treated in this work.

\subsection{Arnoldi iteration}

The Arnoldi iteration \cite{arnoldi} provides the construction of an orthonormal basis for Krylov spaces of open quantum systems \cite{Krylov_arnoldi}: from an operator $\hat u$, it generates an orthonormal set $\{\hat v_0, \dots, \hat v_k, \dots\}$ from the application of the Lindbladian $\mathcal{L}^{\dagger}$ on $\hat u$, such that 
\begin{equation}
    \text{span}(\hat v_0, \dots, \hat v_k) = \text{span}(\hat u, \dots, \mathcal{L}^{\dagger k} \hat u).
\end{equation}
The algorithm is initialized by defining $\hat v_0 = \hat u / \norm{\hat u}$. Then, is iteratively defined, for $k = 1, 2, \dots$:
\begin{itemize}
    \item $\hat u_k = \mathcal{L}^{\dagger} \hat v_{k-1}$, the result of the action of the Lindbladian $\mathcal L^{\dagger}$ onto the Arnoldi operator $\hat v_{k-1}$;
    \item $\tilde u_k = \hat u_k - \sum_{n=0}^{k-1} \langle \hat v_n, \hat u_k\rangle \hat v_n$, the projection of $\hat u_k$ on the complement subspace of $\text{span}(\hat v_0, \dots, \hat v_{k-1})$;
    \item If $\norm{\tilde u_k}>0$, then $\hat v_k = \tilde u_k/\norm{\tilde u_k}$. Otherwise, stop the iteration.
\end{itemize}
For a quantum system with finite dimension, the iteration stops at some index $k=K$ that corresponds to the dimension of the Krylov space of $\hat u$. \\

For the transport-based QST, the Arnoldi iteration provides a practical tool for constructing states from transport measurements. Given a Lindbladian $\mathcal{L}$, we numerically compute for each $P\subseteq\{1, \dots, M\}$, the Arnoldi basis $\{\hat v_0^{P}, \dots, \hat v_{K_P-1}^P\}$ of the Krylov subspace $\mathcal{K}_{P}=\operatorname{span}\bigl\{\mathcal{L}^{\dagger k}\,\hat n_{P}\,\bigl|\,k\in\mathbb{N}\bigr\}$. Because of overlaps between the different Krylov subspaces, the sum of these subspaces, $\mathcal{K} = \sum_P \mathcal{K}_P$, has its dimension $K_{tot}\leq\sum_P K_P$. To perform QST over the states in $\mathcal{K}$, one has to choose an ensemble of indexes $\bigl\{L_P\leq K_P\,\bigl|\,P\subseteq\{1, \dots, M\}\bigr\}$ such that 
\begin{equation}
    \sum_P L_P = K \quad\text{ and } \quad
    \text{span}(\hat v_k^{P}, \, P\subseteq\{1, \dots, M\}, \, k\leq L_P\}) = \mathcal{K}.
\end{equation}
Then measuring the $I_P^{(k)}$ for $P\subseteq\{1, \dots, M\}$ and $k\leq L_P$ determines the parts of the system's state in $\mathcal{K}$ in an optimal way. \\

\section{Appendix B: QST Completeness and Lindbladian spectrum}

In this section, we analyze 
how symmetries, and more generally spectral properties of the Lindblad superoperator, affect the completeness of the transport-based QST
of a general qubit system.
We note $\lambda_i$, $\hat \rho_i$ and $\hat \sigma_i$ respectively the eigenvalues, the right and left eigen-operators of the Linbladian, $\mathcal{L} \hat \rho_i = \lambda_i \hat \rho_i$ and $\mathcal{L}^{\dagger} \hat \sigma_i = \lambda_i^{\ast} \hat \sigma_i$.
For a non-hermitian Lindbladian, the eigen-operators are not in general orthonormal, $\langle \hat \rho_i, \hat \rho_j \rangle \neq \delta_{ij}$. Still, the left and right eigen-operators are bi-orthogonal with respect to the Hilbert-Schmidt product, $\langle \hat \sigma_i, \hat \rho_j \rangle = \Tr[\hat \sigma_i^{\dagger} \hat \rho_j] = \delta_{ij}$. \\

If the Lindbladian is diagonalisable, the left and right eigen-operators each form a complete basis for the Hilbert space. The Krylov operators $\mathcal L^{\dagger k} \hat n_P$ are then set by:
\begin{equation}
	\mathcal L^{\dagger k} \hat n_P = \sum_{i=1}^n \lambda_i^{\ast k} \langle  \hat \rho_i, \hat n_P \rangle \hat \sigma_i,
\end{equation}
with $n$ the dimension of the operator Hilbert space.
The corresponding set of equations for $k=0$ to $n-1$ when evaluating on the state $\densmat(t)$ can be combined into the matrix equation
\begin{equation}
    \vec I = \Lambda^{\ast} V \vec S,
\end{equation}
with $\vec I$ the vector of state coefficients that can be obtained from current measurements,
$$\vec I=(\Tr[\mathcal L^{\dagger k} \hat n_P \densmat(t)])_{0\leq k \leq n-1} = (p_{k,P})_{0\leq k \leq n-1},$$ 
$\Lambda$ the Vandermonde matrix of the Lindbladian eigenvalues,
$$\Lambda = \left (
\begin{smallmatrix}
1 & 1 & \dots & 1 \\
\lambda_1 & \lambda_2 & \dots & \lambda_n \\
\vdots & \vdots & & \vdots \\
\lambda_1^{n-1} & \lambda_2^{n-1} & \dots & \lambda_n^{n-1} 
\end{smallmatrix} \right ),$$
$V$ the diagonal matrix of the overlaps between the occupation number operator $\hat n_P$ and the eigen-operators $\hat \rho_i$,
$$V = \text{Diagonal}(\langle  \hat \rho_i, \hat n_P \rangle, 1\leq i \leq n),$$
and $\vec S$ the coefficients of $\densmat(t)$ in the eigenbasis $\{\hat \rho_i\}$,
$$\vec S = (\Tr[ \hat \sigma_i \densmat])_{1\leq i \leq n}.$$
Note that $\Lambda$ is invertible if and only if the $\lambda_i$'s are all different, and $V$ if the elements $\langle  \hat \rho_i, \hat n_P \rangle$ are non-zero.

In the case of all eigenvalues $\lambda_i$ of the Lindbladian being different, the latter decomposition shows that the Krylov space $\mathcal K_P$ generated by $\hat{n}_P$ contains exactly the left eigen-operators $\hat{\sigma}_i$ for which $\langle \hat{\rho}_i, \hat{n}_P \rangle \neq 0$.
This implies that we may have complete QST when every right eigen-operator $\hat{\rho}_i$ has non-vanishing overlap with at least one occupation-number operator, i.e. $\mathcal K = \text{span}(\hat \sigma_i \vert \exists P\subseteq \{1,\dots,M\}, \langle \hat \rho_i, \hat n_P \rangle \neq 0)$.

If the Linbladian’s spectrum $\mathcal S(\mathcal L)$ has some degeneracies, i.e. eigenvalues that correspond to several eigenvectors, then the Krylov space of $\hat n_P$ changes for $\mathcal K_P = \text{span}(\sum_{\lambda_i = \lambda}  \langle \hat \rho_i, \hat n_P \rangle \hat \sigma_i \vert \lambda \in \mathcal S(\mathcal L))$. \emph{The number of vectors in $\mathcal K_P$ has reduced compared to the case of no degeneracy in the Lindabladian’s spectrum, reducing as much the amount of states that can be determined from the measurements of the $I_P^{(k)}$, $k\in\mathbb N$.} This loss could however be recovered overall by considering several subsets $P_s$, such that the $ \sum_{\lambda_i = \lambda}  \langle \hat \rho_i, \hat n_{P_s} \rangle \hat \sigma_i$ would be linearly independent. \\

If the Linbladian is non-diagonalizable, i.e. in the case of an exceptional point~\cite{Khandelwal2021signatures,Khandelwal2024emergent}, some eigenvalues and their eigen-operators both coalesce. The Lindbladian can then be written in a Jordan canonical form. For example, in the case of an exceptional point of order 2 for the eigenvalue $\lambda_{EP}$, we can define through a Jordan chain the normalized right eigen-operators $\hat \rho_{EP}^{(1)}$ and $\hat \rho_{EP}^{(2)}$ such that  
$(\mathcal L - \lambda_{EP} ) \hat \rho_{EP}^{(1)} = 0 $ and
$(\mathcal L - \lambda_{EP}) \hat \rho_{EP}^{(2)} = A \hat \rho_{EP}^{(1)}$ with $A\in\mathbb C^{\ast}$. The left eigen-operators $\hat \sigma_i$ and $\hat \sigma_{EP}^{(r)}$ are constructed biorthogonal with the right eigen-operators.
The Krylov operators $\mathcal L^{\dagger k} \hat n_P$ then read:
\begin{align}
    \mathcal L ^{\dagger k} \hat n_P =& \sum \lambda_i^{\ast k} c_i^P \hat \sigma_i
    + \lambda_{EP}^{\ast k }c_{EP}^{(2)} \hat \sigma_{EP}^{(2)} + (\lambda_{EP}^{\ast k} c_{EP}^{P(1)} + k \lambda_{EP}^{\ast k-1} A^{\ast}  c_{EP}^{P(2)}) \hat \sigma_{EP}^{(1)}
\end{align}
with $c_i^P = \langle  \hat \rho_i, \hat n_P \rangle$ and $c_{EP}^{P(r)} = \langle  \hat \rho_{EP}^{(r)}, \hat n_P \rangle$ for $r=1,2$. Assuming that the eigenvalues other than $\lambda_{EP}$ have no degeneracy, the Krylov space $\mathcal K_P$ exactly contains the eigenvectors $\hat \sigma_i$ such that $c_i^P\neq 0$, plus $\hat \sigma_{EP}^{(1)}$ (respectively $\hat \sigma_{EP}^{(2)}$) if $c_{EP}^{P(1)}\neq 0$ or $c_{EP}^{P(2)}\neq 0$ (resp. if $c_{EP}^{P(2)}\neq 0$). \emph{Compared to the case of a diagonalizable Lindbladian, we have increased the completeness of the transport-based QST since the condition on measuring the state $\hat \rho_{EP}^{(1)}$ has been loosened while those for other eigen-operators have remain identical.} \\

\section{Appendix C: Study of the two-qubit system}
\label{sct:TQS}

\subsection{Vectorization of the open quantum system}

To gain a more complete understanding of the results presented in the main text on the two-qubit case, we perform row-wise vectorization \cite{vectorization}, which acts on the density matrix written in the Fock basis as
\begin{equation}
    \densmat = \sum p_{n_Ln_R}^{m_Lm_R}\ket{n_L n_R}\bra{m_L m_R} \longrightarrow \lvert \rho \rangle\!\rangle = \sum p_{n_Ln_R}^{m_Lm_R} \ket{m_L m_R} \otimes \ket{n_L n_R},
\end{equation}
and transforms superoperators into operators as
\begin{equation}
    \mathcal{A} \densmat = \hat A_1 \densmat \hat A_2 \longrightarrow \mathbf{A} = (\hat{A}_1 \otimes \hat{A}_2^T) \lvert \rho \rangle\!\rangle.
\end{equation}
This representation is specially convenient for analytics and numerics, as the quantum objects can be easily created using Kroenecker products. \\

For the sake of our analysis, we hereafter move to the basis where the density matrix
\begin{equation} \label{eq:densmat}
    \rho = \begin{pmatrix}
r_{00} & v & x & \beta \\
v^{\ast} & r_{01} & \alpha & y \\
x^{\ast} & \alpha^{\ast} & r_{10} & z \\
\beta^{\ast} & y^{\ast} & z^{\ast} & r_{11} \\
\end{pmatrix}
\end{equation}
is described by the vector
\begin{equation}
    \vec \rho = (r_{00}, r_{01}, r_{10}, r_{11}, \aI, \aR, \bI, \bR, \I{x}, \R{x}, \I{y}, \R{y}, \I{v}, \R{v}, \I{z}, \R{z})^T.
\end{equation}

\subsection{Structure of the Lindbladian}

Following this vectorization, the evolution equation resumes to a linear differential equation:
\begin{equation}
    \partial_t \vec \rho(t) = \mathbf{L} \vec \rho(t)
\end{equation}
with the Lindblad matrix $\mathbf{L} = \mathbf{L}_\text{unit} + \mathbf{L}_\text{diss}$ made of two components: a unitary-evolution matrix
\begin{equation}
    \mathbf{L}_\text{unit} = - i (H_S \otimes \mathbb{I}_4 - \mathbb{I}_4 \otimes H_S^T)
\end{equation}
and a dissipative-evolution matrix $\mathbf{L}_\text{diss} = \sum_{j=L,R} (\mathbf{L}_j^++\mathbf{L}_j^-+\mathbf{L}_j^z)$, with
\begin{align}
    \mathbf{L}_j^s = \gamma_j^s( \sigma_s^{(j)} \otimes \sigma_s^{(j)} 
    - 
    \frac{1}{2}(\sigma_s^{(j)\dagger} \sigma_s^{(j)} \otimes \mathbb{I}_4 + \mathbb{I}_4 \otimes (\sigma_s^{(j)\dagger} \sigma_s^{(j)})^T))
\end{align}
for $s = R,L,z$. The matrices $\mathbf{L}_\text{unit}$ and $\mathbf{L}_\text{diss}$ may be decomposed into 16 blocks of dimension $4\times 4$, corresponding to the different components of the dynamics:
\begin{enumerate}[label=\roman*)]
\item The evolution set by the dissipative terms, induced by the presence of reservoirs and of pure dephasing, is captured by the block-diagonal matrix
\begin{equation}
    \mathbf{L}_{\text{diss}} = 
\begin{pmatrix}
     \mathbf{L}_{\text{diss}}^{\text{pop}} & & & \\
     & \mathbf{L}_{\text{diss}}^{\alpha \beta} & & \\
     & & \mathbf{L}_{\text{diss}}^{x y} & \\
     & & & \mathbf{L}_{\text{diss}}^{vz}
\end{pmatrix}
\end{equation}
with the subspace matrices
\begin{eqnarray}
    &\mathbf{L}_{\text{diss}}^{\text{pop}} = \left (\begin{smallmatrix}
     - (\gLp + \gRp) & \gRm & \gLm & 0 \\
     \gRp & -(\gLp + \gRm) & 0 & \gLm \\
     \gLp & 0 & - (\gLm + \gRp) & \gRm \\
     0 & \gLp & \gRp & - (\gLm + \gRm)
 \end{smallmatrix} \right ) \qquad 
 &\mathbf{L}_{\text{diss}}^{xy} = -\frac{\GL + 2 \gamma_L^z}{2} \mathbb{I}_4 +
 \left ( \begin{smallmatrix}
     -\gRp & 0 & \gRm & 0 \\
     0 & -\gRp & 0 & \gRm \\
     \gRp & 0 & -\gRm & 0 \\
     0 & \gRp & 0 & -\gRm
 \end{smallmatrix} \right ) \\
 &\mathbf{L}_{\text{diss}}^{\alpha \beta} = -\frac{\tilde \Gamma}{2} \mathbb{I}_4  \qquad
 &\mathbf{L}_{\text{diss}}^{vz} = -\frac{\GR + 2 \gamma_R^z}{2} \mathbb{I}_4 +
 \left ( \begin{smallmatrix}
     -\gLp & 0 & \gLm & 0 \\
     0 & -\gLp & 0 & \gLm \\
     \gLp & 0 & -\gLm & 0 \\
     0 & \gLp & 0 & -\gL
 \end{smallmatrix} \right )
\end{eqnarray}
Here, $\Gamma_r=\gamma_r^+ + \gamma_r^-$ for $r=L,R$, $\Gamma = \GL + \GR$, $\Gamma_z=\gamma_L^z + \gamma_R^z$ and $\tilde \Gamma = \Gamma + \Gamma_z$. $\mathbf{L}_{\text{diss}}$ exponentially suppresses the coherences, as expected, while generating coupling between the populations and between pairs of coherences, $(x, y)$ and $(v, z)$. 

\item On-site energies and interaction, $\hat H_\text{on-site}=\epsilon_L \hat{n}_L + \epsilon_R \hat{n}_R + U \hat{n}_L \hat{n}_R$, yield a superoperator which is also block diagonal in this basis: 
\begin{equation}
    \mathbf{L}_{\text{on-site}} = 
\begin{pmatrix}
     \mathbf{L}_{\text{on-site}}^{\text{pop}} & & & \\
     & \mathbf{L}_{\text{on-site}}^{\alpha \beta} & & \\
     & & \mathbf{L}_{\text{on-site}}^{x y} & \\
     & & & \mathbf{L}_{\text{on-site}}^{vz}
\end{pmatrix}
\end{equation}
with the subspace matrices
\begin{eqnarray}
     &\mathbf{L}_{\text{on-site}}^{\text{pop}} = 0 \qquad
     &\mathbf{L}_{\text{on-site}}^{xy} = 
 \left (\begin{smallmatrix}
     0 & \epsilon_L &  &  \\
     -\epsilon_L & 0 &  &  \\
      &  & 0 & \epsilon_L + U \\
      &  & -(\epsilon_L + U) & 0
 \end{smallmatrix} \right ) \\
 &\mathbf{L}_{\text{on-site}}^{\alpha \beta} = 
 \left (\begin{smallmatrix}
     0 & \delta &  &  \\
     -\delta & 0 &  &  \\
      &  & 0 & E \\
      &  & -E & 0
 \end{smallmatrix} \right )  \qquad
 &\mathbf{L}_{\text{on-site}}^{vz} = 
 \left (\begin{smallmatrix}
     0 & \epsilon_R &  &  \\
     -\epsilon_R & 0 &  &  \\
      &  & 0 & \epsilon_R + U \\
      &  & -(\epsilon_R + U) & 0
 \end{smallmatrix} \right )
\end{eqnarray}

It couples real and imaginary parts of the coherences.

\item The other interactions between the two qubits, set by the Hamiltonian $\hat H_{\text{int}} = g_\text{res} \, (\hat\sigma_+^{(L)} \hat\sigma_-^{(R)} + \hat\sigma_-^{(L)} \hat\sigma_+^{(R)}) + g_\text{off} \, (\hat\sigma_+^{(L)} \hat\sigma_+^{(R)} + \hat\sigma_-^{(L)} \hat\sigma_-^{(R)})$, correspond to the superoperator:
\begin{equation}
    \mathbf{L}_{\text{on-site}} = 
\begin{pmatrix}
     0 & -2\mathbf{L}_{\text{int}}^{(1)T} & 0 & 0 \\
     \mathbf{L}_{\text{int}}^{(1)} & 0 & 0 & 0 \\
     0 & 0 & 0 & -\mathbf{L}_{\text{int}}^{(2)T} \\
     0 & 0 & \mathbf{L}_{\text{int}}^{(2)} & 0
\end{pmatrix}
\end{equation}
with
\begin{align}
\mathbf{L}_{\text{int}}^{(1)} &= 
 \left (\begin{smallmatrix}
     0 & g_\text{res} & -g_\text{res} & 0 \\
     0 & 0 & 0 & 0 \\
     0 & 0 & 0 & 0 \\
     0 & 0 & 0 & 0
 \end{smallmatrix} \right )
+\left (\begin{smallmatrix}
     0 & 0 & 0 & 0 \\
     0 & 0 & 0 & 0 \\
     g_\text{off} & 0 & 0 & -g_\text{off} \\
     0 & 0 & 0 & 0
 \end{smallmatrix} \right ), \\
\mathbf{L}_{\text{int}}^{(2)}&= 
 \left ( \begin{smallmatrix}
     0 & g_\text{res} & 0 & 0 \\
     -g_\text{res} &  & 0 & 0 \\
     0 & 0 &  & -g_\text{res} \\
     0 & 0 & g_\text{res} & 0
 \end{smallmatrix} \right )
+\left ( \begin{smallmatrix}
     0 & 0 & 0 & -g_\text{off} \\
     0 & 0 & -g_\text{off} & 0 \\
     0 & g_\text{off} & 0 & 0 \\
     g_\text{off} & 0 & 0 & 0
 \end{smallmatrix} \right ).
\end{align}
The resonant part of the inter-qubit interaction couples the populations and the dynamics of the first off-diagonal coherence, $\alpha(t)$, while the off-resonant part couples the populations and the dynamics of the second off-diagonal coherence, $\beta(t)$. They also interconnect the $(x,y)$ and $(v,z)$ pairs of coherences.
\end{enumerate}

\subsection{Generalization towards complete QST} 
\label{sct:generalization}

To perform complete QST of a full density matrix, one needs to consider a Hamiltonian with additional elements - for example, local drives on the qubits, $\hat H_\text{drive} := f_L (\sLp+\sLm) + f_R (\sRp+\sRm)$ with $f_L$ and $f_R$ real amplitudes of the drives. The matrix of the associated superoperator, decomposed into $4\times4$ blocks, reads:
\begin{equation}
    \mathbf{L}_{\text{drive}} = 
\begin{pmatrix}
    0 & 0 & -2\mathbf{L}_{\text{drive}}^{(L1)T} & 0 \\
    0 & 0 & 0 & -\mathbf{L}_{\text{drive}}^{(L2)T} \\
    \mathbf{L}_{\text{drive}}^{(L1)} & 0 & 0 & 0 \\
    0 & \mathbf{L}_{\text{drive}}^{(L2)} & 0 & 0
\end{pmatrix}
+ 
\begin{pmatrix}
    0 & 0 & 0 & -2\mathbf{L}_{\text{drive}}^{(R1)T} \\
    0 & 0 & -\mathbf{L}_{\text{drive}}^{(R2)T} & 0 \\
    0 & \mathbf{L}_{\text{drive}}^{(R2)} & 0 & 0 \\
    \mathbf{L}_{\text{drive}}^{(R1)} & 0 & 0 & 0
\end{pmatrix}
\end{equation}
with
\begin{eqnarray}
&\mathbf{L}_{\text{drive}}^{(L1)} = 
 \left (\begin{smallmatrix}
     f_L & 0 & -f_L & 0 \\
     0 & 0 & 0 & 0 \\
     0 & f_L & 0 & -f_L \\
     0 & 0 & 0 & 0
 \end{smallmatrix} \right ) \qquad
 &\mathbf{L}_{\text{drive}}^{(L2)} = 
 \left ( \begin{smallmatrix}
     0 & -f_L & 0 & f_L \\
     -f_L &  & -f_L & 0 \\
     0 & f_L &  & -f_L \\
     f_L & 0 & f_L & 0
 \end{smallmatrix} \right ) \\
&\mathbf{L}_{\text{drive}}^{(R1)} = 
 \left (\begin{smallmatrix}
     f_R & -f_R & 0 & 0 \\
     0 & 0 & 0 & 0 \\
     0 & 0 & f_R & -f_R \\
     0 & 0 & 0 & 0
 \end{smallmatrix} \right ) \qquad
 &\mathbf{L}_{\text{drive}}^{(R2)} = 
 \left ( \begin{smallmatrix}
     0 & -f_R & 0 & f_R \\
     f_R &  & -f_R & 0 \\
     0 & f_R &  & -f_R \\
     -f_R & 0 & f_R & 0
 \end{smallmatrix} \right )
\end{eqnarray}
The drive on qubit $L$ (respectively on qubit $R$) couples the evolution of the populations and with the one of the $(x,y)$ (resp. $(v,z)$) coherence pair of coherences, and of $(\alpha, \beta)$ and $(v,z)$ (resp. $(x,y)$). 

The resulting total Hamiltonian $\hat H_\text{tot} = \hat H_\text{on-site} + \hat H_\text{int} + \hat H_\text{drive}$ is a full matrix:
\begin{equation}
    \hat H_\text{tot} = \begin{pmatrix}
0 & f_R & f_L & g_\text{off} \\
f_R  & \epsilon_R & g_\text{res}  & f_L \\
f_L  & g_\text{res}  & \epsilon_R  & f_R  \\
g_\text{off}  & f_L  & f_R  &  E \\
\end{pmatrix}
\end{equation}
with $E= \epsilon_R + \epsilon_L + U$ the energy of the doubly-occupied state.
In this case, all coherences can be constructed from Krylov states of the occupation-number operators, and the transport-based QST is informally complete. For example, the first derivatives of the occupation numbers read:
\begin{align}
    \dot n_L(t) &= \gLp - \GL n_L + 2 g_{\text{res}} \I{\alpha(t)} + 2 g_{\text{off}} \I{\beta(t)} + f_L\I{x(t)+y(t)}, \\
    \dot n_R(t) &= \gRp - \GR n_R - 2 g_{\text{res}} \I{\alpha(t)} + 2 g_{\text{off}} \I{\beta(t)} +  f_R \I{v(t)+z(t)}.
\end{align}
The imaginary part of the $\alpha$ and $\beta$ coherences cannot be determined from $(\dot I_L, I_L, \dot I_R, I_R)$ anymore. To get a closed set of equations on the coherences, one needs to consider additional equations, obtained similarly from higher time-derivatives. \\

\section{Appendix D: Current correlation functions}
\label{sct:current_corr}

\subsection{Definitions}

The quantum jumps of excitations tunneling into (respectively out of) the system, induced by the presence of reservoir $j$, are described within the local Lindblad master equation formalism by the superoperators
\begin{align}
    \mathcal{L}_j^+ &:= \gamma_j^+ \sigma_+^{(j)} \bullet \sigma_-^{(j)}, \\
    \mathcal{L}_j^- &:= \gamma_j^- \sigma_-^{(j)} \bullet \sigma_+^{(j)}.
\end{align}
It was recently derived in \cite{Blasi2024}, using a full-counting statistics approach, that within this formalism, the correlation function $S_{j_1j_2}(t_1,t_2)$ between the currents in reservoir $j_1$ at time $t_1$ and $j_2$ at time $t_2$ equals to
\begin{align}
    S_{j_1j_2}(t_1,t_2) := 
    & \delta_{j_1, j_2}\delta(t_1-t_2)\Tr{\mathcal{A}_{j_1}\densmat(t_1)} 
    \notag \\
    &+ \Theta(t_1-t_2) \Tr{\mathcal{I}_{j_1} \e{\mathcal{L}(t_1-t_2)} \mathcal{I}_{j_2}\densmat(t_2)} \notag \\
    &+ \Theta(t_2-t_1) \Tr{\mathcal{I}_{j_2}\e{\mathcal{L}(t_2-t_1)}\mathcal{I}_{j_1}\densmat(t_1)}
    \notag \\
    &-\Tr{\mathcal{I}_{j_1}\densmat(t_1)} \Tr{\mathcal{I}_{j_2}\densmat(t_2)},
\end{align}
with $\mathcal{I}_j:=\mathcal{L}_j^+ - \mathcal{L}_j^-$ and $\mathcal{A}_j:=\mathcal{L}_j^+ + \mathcal{L}_j^-$ the current and activity superoperators associated to reservoir $j$. The first term of this expression, the dynamical activity $A_{j_1}(t)= \Tr[\mathcal{A}_{j_1}\densmat(t)]$, is present only for instantaneous auto-correlation functions and describes the rate of jumps occurring at the interface with the reservoir $j$, regardless of their direction~\cite{Blasi2024,Bourgeois2024,Blasi2025}. 

\subsection{Instantaneous auto-correlation functions}

The instantaneous ($t_1=t_2=t$) auto-correlation functions are formally given by:
\begin{equation}
    S_{j,j}(t,t) = \delta(0) A_j(t) + \Tr[\mathcal{I}_j^2 \densmat(t)] - I_j(t)^2.
\end{equation}
The presence of the $\delta$-distribution in this definition calls for a special care when manipulating $S_{j,j}(t,t)$. This is usually done by working in the frequency domain, analyzing measurements performed over a time window rather than at a fixed time. As for the transport-based QST, we note that the instantaneous auto-correlation functions do not provide further information on the quantum state than the current averages, since
\begin{align}
    \Tr[\mathcal{I}_j^2 \densmat(t)] - I_j(t)^2 &= -\Gamma_j (\gamma_j^+ (1-n_j(t))^2 + \gamma_j^- n_j(t)^2), \\
    A_j(t) &= \gamma_j^+ (1-n_j(t)) + \gamma_j^- n_j(t).
\end{align}
For this reason, we believe the instantaneous auto-correlation functions would not be relevant for transport-based QST.

\subsection{Two-time correlation functions}

In general, the superoperator $\e{\mathcal{L}(t_1-t_2)}$ corresponds to a very complex matrix, whose analytical expression is not to be given here. Still, for two-qubit systems, one may explore the information contained in the two-time correlation functions by studying the matrices
\begin{align*}
    \mathcal{I}_L \rho(t) = \begin{pmatrix}
-\gLm r_{10}(t) & -\gLm z(t) & 0 & 0 \\
-\gLm z^{\ast}(t) & -\gLm r_{11}(t) & 0 & 0 \\
0 & 0 & \gLp r_{00}(t) & \gLp v(t) \\
0 & 0 & \gLp v^{\ast}(t) & \gLp r_{01}(t) \\
\end{pmatrix}  \\
    \mathcal{I}_R \rho(t) = \begin{pmatrix}
-\gRm r_{01}(t) & 0 & - \gRm y(t) & 0 \\
0 & \gRp r_{00}(t) & 0 & \gRp x(t) \\
- \gRm y^{\ast}(t) & 0 & -\gRp r_{11}(t) & 0 \\
0 & \gRp x^{\ast}(t) & 0 & \gRp r_{10}(t) \\
\end{pmatrix}
\end{align*}
where $r_{ij}$, $v$, $x$, $y$ and $z$ are elements of the density matrix $\rho$, as introduced in Eq.~\eqref{eq:densmat}.

Importantly, the coherences $\alpha(t)$ and $\beta(t)$ are absent from these matrices, 
yielding that no information about these coherences can be obtained from $S_{j_1j_2}(t_1,t_2)$ itself.
This aspect justifies the need for other transport quantities, like current-average time-derivatives, to perform complete transport-based QST.

\section{Appendix E: How to access the system's parameters}
\label{app:dynamics}

\subsection{Main result}

The QST protocol introduced in this work relies on the precise knowledge of the system dynamics parameters.  If the system's Hamiltonian $\hat{H}_S$ or the rates of the dissipation processes are not known quantitatively, it is first necessary to measure them. An extension of the transport-based QST protocol precisely allows one to estimate these dynamics parameters from additional transport observables, higher-order time derivatives of the current averages. In Table \ref{tab:evl}, we summarize all necessary transport quantities to be measured to achieve such determination for the two-qubit system presented in the main text, depending on the completeness of $\hat H_S$. 

\begin{table}[h]
\begin{NiceTabular}{c||c|c}[corners, hlines]
Condition on $ \hat H_S$   
& Dynamics parameters 
& Transport quantities
\\ \Hline \Hline

& \Block[c]{}{\mygreen{$g_\text{res}$}, \mygreen{$g_\text{off}$},\\ \mygreen{$\delta$}, \mygreen{$E$}, \\ \mygreen{$\Gamma_z$}}
& \Block[c]{}{\mygreen{$I_L^{(0)}(t_i),...,I_L^{(3)}(t_i)$}\\ and\\ \mygreen{$I_R^{(0)}(t_i),...,I_R^{(3)}(t_i)$}\\ for $i=1,...,5$}
\\ 

\Block[c]{}{$\delta=0$\\ $E=0$} 
& \Block[c]{}{\mygreen{$g_\text{res}$}, \mygreen{$g_\text{off}$}}
& \Block[c]{}{\mygreen{$I_L^{(0)}(t_i),...,I_L^{(2)}(t_i)$}\\ and\\ \mygreen{$I_R^{(0)}(t_i),...,I_R^{(2)}(t_i)$}\\ for $i=1,2,3$}
\\ 

$g_\text{off}=0$
& \Block[c]{}{\mygreen{$g_\text{res}$},\\ \mygreen{$\delta$},\\ \mygreen{$\Gamma_z$}}
& \Block[c]{}{\mygreen{$I_L^{(0)}(t_i),...,I_L^{(3)}(t_i)$}\\ or\\ \mygreen{$I_R^{(0)}(t_i),...,I_R^{(3)}(t_i)$}\\ for $i=1,2,3$}
\\ 
\end{NiceTabular}
\caption{Summary of the transport-based dynamics probing, when all Hamiltonian and pure-dephasing parameters are missing experimental data.}
\label{tab:evl}
\end{table}

\subsection{Formal argument} 

The operators $\mathcal{L}^{\dagger k} \hat n_P$, for $k\in\mathbb N$ and $P\subseteq\{1,\dots,M\}$, are not all linearly independent. A set of current moments $I_P^{(k)}$, with several pairs $(k, P)$, can thus contain redundant information on the system's state, that can be used to probe the evolution parameters.

For example considering a single subset $P$, the Krylov space generated from $\hat n_P$ has a finite dimension that we denote $K_P$: $\mathcal K_P = \text{span}(\hat n_P, \mathcal L^{\dagger} \hat n_P,\dots,\mathcal{L}^{\dagger K_P-1} \hat n_P)$. We thus have $\mathcal{L}^{\dagger K_P}\hat n_P = \sum_{k=0}^{K_P-1} c_k^P \mathcal{L}^{\dagger k}\hat n_P$ with $c_k^P \equiv \langle \mathcal{L}^{\dagger k}\hat n_P, \mathcal{L}^{\dagger K_P}\hat n_P \rangle=\Tr[\hat n_P \mathcal L ^{\dagger (K_P-k)} \hat n_P] = \Tr[\mathcal L ^{\dagger (K_P-k)} \hat n_P]$ being functions of the evolution parameters. Then measuring the current quantities $I_P(t), I_P'(t), \dots I_P^{(K_P)}(t)$ provides the following equation on the $c_k^P$'s:
\begin{equation}
    \sum_{k=0}^{K_P} c_k^P p_{P,k}(t) = 0,
\end{equation}
with $c_{K_P}^P\equiv-1$, and the $p_{P,k}(t)$ given by the general QST expression in the main text. Measuring $I_P(t_i), I_P'(t_i), \dots I_P^{(K_P)}(t_i)$ for different subsets $P$ and different times $t_i$ thus builds a set of conditions on the $c_k^P$'s that can be used to determine some evolution parameters. \\

\subsection{Illustration with the two-qubit system}

We illustrate this method with a two-qubit system.

In the simple case of two degenerate qubits, $\epsilon_L = \epsilon_R \equiv \epsilon$, and a resonant-only interaction, $g_\text{off} = 0$, the dynamics only couples the population $r_{00}, r_{01}, r_{10}, r_{11}$ to the imaginary coherence $i \aI $. The transport-based QST's set of equations then resume to
\begin{subequations}
\label{eq:populations}
\begin{align}
    r_{01}(t) &= -\frac{S_{LR}(t)}{\GL \GR} - \frac{I_R(t)-\gR^+}{\GR} \frac{I_L(t) + \gamma_L^-}{\Gamma}, \\
    r_{10}(t) &= - \frac{S_{LR}(t)}{\GL \GR} - \frac{I_L(t)-\gL^+}{\GL} \frac{I_R(t) + \gamma_R^-}{\Gamma}, \\
    r_{11}(t) &= \frac{S_{LR}(t)}{\GL \GR} + \frac{I_L-\gL^+}{\GL} \frac{I_R-\gR^+}{\GR}, \\
    \aI &= \frac{1}{4 \, g_\text{res}} \Big ( - \frac{\Dot{I}_L}{\GL} - I_L + \frac{\Dot{I}_R}{\GR} + I_R \Big) \,. \label{eq:aI}
\end{align}
\end{subequations}
The transport QST expression for $\bI$ is transformed in this case into the current conservation identity $\frac{\dot{I}_L}{\GL} + I_L + \frac{\dot{I}_R}{\GR} + I_R=0$. Hence, the transport quantities needed for performing QST are simply $S_{LR}(t)$, $I_L(t)$, $I_R(t)$ and $\dot{I}_L(t)$. Further differentiating in time in Eq.~\eqref{eq:aI} yields
\begin{align}
    &\frac{\Ddot{I}_L}{\GL} + \dot{I}_L + \frac{\Tilde{\Gamma}}{2}(\frac{\dot{I}_L}{\GL} + I_L) +  2g_\text{res}^2(\frac{I_L-\gLp}{\GL} - \frac{I_R-\gRp}{\GR}) = 0.
\end{align}
If $g_\text{res}$ is known \emph{a priori}, the additional measurement of $\ddot{I}_L(t)$ at time $t$ provides the value of $\tilde \Gamma$, and thus of the pure dephasing strength $\Gamma_z$. Otherwise, measuring $\Ddot{I}_L$, $\dot{I}_L$, $I_L$ and $I_R$ at two different times $t_1$, $t_2$ determines both $g_\text{res}$ and $\tilde \Gamma$. For error-prone measurements, the number of probe times may be increased to improve the fidelity of this protocol.

In the more general case studied in the main text, differentiating in time in the transport QST expressions for $\aR$ and $\bR$ provides the equations:
\begin{subequations}
\begin{align}
    0 =& \, \Delta \ddot \varphi(t_i)
    + \tilde \Gamma \, \Delta \dot \varphi(t_i) 
    + (\frac{\tilde \Gamma^2}{4} + \delta^2) \, \Delta \varphi(t_i) 
    + 4g_\text{res}^2 \, \Delta \dot \chi(t_i) 
    + 2g_\text{res}^2 \tilde\Gamma \, \Delta \chi(t_i), \\
    0 =&  \; \;  \ddot \Phi(t_i) \;
    + \; \tilde \Gamma \;  \dot \Phi(t_i) \;
    + \; (\frac{\tilde \Gamma^2}{4} + E^2)  \; \Phi(t_i) \;
    + \; 4 \, g_\text{off}^2  \; \dot X(t_i) 
    + \; 2 \, g_\text{off}^2 \tilde \Gamma \; X(t_i),
\end{align}
\end{subequations}
with
\begin{eqnarray*}
    & \,\, \Delta \varphi(t) = \varphi_L(t) - \varphi_R(t), \quad 
    & \, \Delta \chi(t) = \chi_L(t) -\chi_R(t), \\ 
    & \,\, \;\;\; \Phi(t) = \varphi_L(t) + \varphi_R(t), \quad 
    & \, \;\; X(t) = \chi_L(t) + \chi_R(t), \\
    & \varphi_j(t) = \frac{\dot I_j(t)}{\Gamma_j} + I_j(t), \quad
    & \; \; \chi_j(t) = \frac{I_j(t)-\gamma_j^+}{\Gamma_j}.
\end{eqnarray*} 
In the case of an experimental set-up where all the Hamiltonian parameters are known, measuring $\dddot I_j(t)$  ($j=L$ or $R$), in addition to $\ddot I_j(t)$, $\dot I_L(t)$, $\dot I_R(t)$, $I_L(t)$ and $I_R(t)$, determines the pure dephasing strength $\Gamma_z = \frac{1}{2}(\tilde \Gamma- \Gamma)$. 
Otherwise, first probing the four Hamiltonian parameters $g_\text{res}, g_\text{off}, \delta, E$ requires the measurement of $I_j^{(k)}(t_i)$ for $j=L,R$ and $k=0,1,2,3$ at, at least, four different times $t_i$. \\

\section{Appendix F: Connection with current conservation laws}
\label{app:internal_current}

The tomography technique proposed in this work can be understood as a set of transport identities by introducing new transport quantities, internal to the bipartite system. In this section, we illustrate such link with the two-qubit system studied in the main text.

\subsection{Extra-system transport}

The current flowing from reservoir $j$ to its connected qubit is captured in the Master-Equation formalism through the action of the superoperator $\mathcal{D}_j = \gamma_j^+ D[\hat \sigma_+^{(j)}] + \gamma_j^{-} D[\hat \sigma_-^{(j)}]$, and of the occupation-number operator $\hat n_j$, on the state. One can indeed derive that the qubit-reservoir current satisfies:
\begin{equation}
    I_j(t) = \Tr[\hat{n}_j \mathcal{D}_j\densmat(t)].
\end{equation}

\subsection{Intra-system transport}

The Hamiltonian part $\hat H_\text{res} = g_\text{res} (\sigma_+^{(L)}\sigma_-^{(R)} + \sigma_-^{(L)}\sigma_+^{(R)})$ preserves the total number of particle in the bipartite system,
\begin{equation}
    \Tr([\hat{n}_L + \hat{n}_L, \hat H_\text{res}]\densmat) = 0.
\end{equation}
Its action generates particle transfers between the two qubits, inducing a current from the left to the right party equal, by definition, to
\begin{equation}
    I_S := i \Tr([\hat{n}_L, \hat H_\text{res}]\densmat) = -i \Tr([\hat{n}_R, \hat H_\text{res}]\densmat).
\end{equation}
For any generic density matrix, this internal current is determined by the $\alpha$-coherence as
\begin{equation} \label{eq:def_IS}
    I_S = -2 g_\text{res} \aI.
\end{equation}

Meanwhile, the Hamiltonian term $\hat H_\text{off} = g_\text{off} (\sigma_+^{(L)}\sigma_+^{(R)} + \sigma_-^{(L)}\sigma_-^{(R)})$ preserves the particle imbalance between the two parties,
\begin{equation}
    \Tr([\hat{n}_L - \hat{n}_L, \hat H_\text{off}]\densmat) = 0.
\end{equation}
Its action corresponds to a common particle production in the two qubits, given by
\begin{equation} 
    P_S := -i \Tr([\hat{n}_L, \hat H_\text{off}]\densmat) = -i \Tr([\hat{n}_R, \hat H_\text{off}]\densmat).
\end{equation}
For any generic density matrix, this internal current is determined by the $\beta$-coherence as
\begin{equation}
\label{eq:def_PS}
    P_S = 2 g_\text{off} \bI.
\end{equation}

\subsection{Current conservation laws}
The time evolution of the local occupation number $n_j$ for $j=L,R$ is set by
\begin{equation}
    \dot{n}_j = \Tr[\hat n_j \mathcal L\densmat]= \Tr(\hat{n}_j \mathcal{D}_j\densmat) -i \Tr([\hat{n}_j, H_I + H_J]\densmat),
\end{equation} 
Overall, we thus have the following transport identities:
\begin{subequations} \label{eq:transport_id}
\begin{align}
    \dot{n}_L 
    &= I_L - I_S + P_S, \\
    \dot{n}_R 
    &= I_R + I_S + P_S.
\end{align}
\end{subequations}
Substituting $n_j$ by its transport-based expression, one obtains the following equations on the internal and external transport quantities:
\begin{subequations} \label{eq:transport_id_full}
\begin{align}
    I_S &= \frac{1}{2}(\frac{\Dot{I}_L}{\GL} + I_L -\frac{\Dot{I}_R}{\GR} - I_R), \\
    P_S &= - \frac{1}{2}(\frac{\Dot{I}_L}{\GL} + I_L + \frac{\Dot{I}_R}{\GR} + I_R).
\end{align}
\end{subequations}
This identity exactly corresponds to the QST expressions for the imaginary part of the coherences. \\
Similarly, differentiating in time in Eq.~\eqref{eq:def_IS}-\eqref{eq:def_PS}, yields
\begin{subequations} \label{eq:der_IS_PS}
\begin{align}
    \dot{I}_S &= -\frac{\Gamma}{2}I_S - 2g_\text{res} \de \aR - 2g_\text{res}^2 (\frac{I_L-\gLp}{\GL}-\frac{I_R-\gRp}{\GR}), \\
    \dot{P}_S &= -\frac{\Gamma}{2}P_S + 2g_\text{off} E \bR + 2g_\text{off}^2 (1+\frac{I_L-\gLp}{\GL}+\frac{I_R-\gRp}{\GR}),
\end{align}
\end{subequations}
which is a variant of the QST expressions for the real part of the coherences.

In conclusion, the tomography technique proposed in this work is based on the following set of correspondences: the current averages $I_L$ and $I_R$ directly determine the populations $n_L$ and $n_R$; their time derivatives $\dot{I}_L$ and $\dot{I}_R$ provides information on the imaginary coherences $\aI$ and $\bI$ via the internal particle current and production $I_S$ and $P_S$; and finally, the second time derivatives $\Ddot{I}_L$ and $\Ddot{I}_R$ are linked with the the real coherences $\aR$ and $\bR$ thanks to the time-derivation of $I_S$ and $P_S$. \\

In the steady-state, current conservation identities in each qubit lead to the equations:
\begin{subequations} \label{eq:transport_id_ss}
\begin{align}
    I_S^{ss} 
    &= (I_L^{ss} - I_R^{ss})/2, \\
    P_S^{ss}
    &= -(I_R^{ss} + I_R^{ss})/2.
\end{align}
\end{subequations}
The steady-state internal current $I_S^{ss}$ is thus equal to the average current flowing from the left to the right reservoir, while the steady-state internal production $P_S^{ss}$ counterbalances the average release of particles out of the system.

\section{Appendix G: Transport-based entanglement measures}
\label{app:concurrence}

For the two-qubit system presented in the main text, initiated with an X-shaped density matrix, the concurrence is given at all time in terms of transport observables as:
\begin{equation}
    \mathcal{C} = 2 \max\{0, \vert \alpha \vert -
\sqrt{r_{00} r_{11}}, \vert \beta \vert -
\sqrt{r_{01} r_{10}}\}
\end{equation}
where
\begin{align}
    \vert \alpha \vert -
\sqrt{r_{00} r_{11}} =& 
\frac{1}{\abs{2 g_\text{res}}}\sqrt{(\varphi_L-\varphi_R)^2 
+ 
\frac{1}{\de^2} 
(\dot \varphi_L - \dot \varphi_R
+
\frac{\Tilde{\Gamma}}{2}(\varphi_L-\varphi_R) 
+ 
4g_\text{res}^2 (\chi_L - \chi_R))^2} \nonumber \\
& \qquad 
- \sqrt{
    \Big (\frac{S_{LR}}{\GL \GR} + \chi_L \chi_R \Big ) 
    \Big (\frac{S_{LR}}{\GL \GR} + (\chi_L +1) (\chi_R +1) \Big ) }
\end{align}
and
\begin{align}
\vert \beta \vert -
\sqrt{r_{01} r_{10}} =& 
\frac{1}{\abs{2 g_\text{off}}}\sqrt{(\varphi_L+\varphi_R)^2 
+ 
\frac{1}{E^2} 
(\dot \varphi_L + \dot \varphi_R
+
\frac{\Tilde{\Gamma}}{2}(\varphi_L+\varphi_R) 
+ 
4g_\text{off}^2 (\chi_L + \chi_R +1))^2} \nonumber \\
& \qquad 
- \sqrt{
    \Big (\frac{S_{LR}}{\GL \GR}+ \chi_L (\chi_R +1) \Big )
    \Big (\frac{S_{LR}}{\GL \GR} + (\chi_L+1) \chi_R \Big ) 
     }
\end{align}
with $\varphi_j(t) = \frac{\dot I_j(t)}{\Gamma_j} + I_j(t)$ and $\chi_j(t) = \frac{I_j(t)-\gamma_j^+}{\Gamma_j}$. For the sake of clarity, time-dependence is implicit here.